\documentclass[a4paper]{article}
\usepackage[numbers,sort]{natbib}

\usepackage{INTERSPEECH2018}

\usepackage{graphicx}
\usepackage{amsmath}
\usepackage{amsthm}
\usepackage{subcaption}
\usepackage{enumitem}

\usepackage{hyperref}
\hypersetup{hidelinks}

\def\equationautorefname#1#2\null{
	Eq.(#2\null)
}

\title{Parameter-Efficient Conformers via Sharing Sparsely-Gated Experts for End-to-End Speech Recognition}
\name{Ye Bai, Jie Li, Wenjing Han, Hao Ni, Kaituo Xu, Zhuo Zhang, Cheng Yi, Xiaorui Wang}
\address{Kuaishou Technology Co., Ltd, Beijing, China}

\email{\{baiye03,lijie03,hanwenjing,nihao,xukaituo,zhangzhuo03,chengyi03,wangxiaorui\}@kuaishou.com}

\begin{document}

\maketitle

\begin{abstract}
While transformers and their variant conformers show promising performance in speech recognition, the parameterized property leads to much memory cost during training and inference. Some works use cross-layer weight-sharing to reduce the parameters of the model. However, the inevitable loss of capacity harms the model performance. To address this issue, this paper proposes a parameter-efficient conformer via sharing sparsely-gated experts. Specifically, we use sparsely-gated mixture-of-experts (MoE) to extend the capacity of a conformer block without increasing computation. Then, the parameters of the grouped conformer blocks are shared so that the number of parameters is reduced. Next, to ensure the shared blocks with the flexibility of adapting representations at different levels, we design the MoE routers and normalization individually. Moreover, we use knowledge distillation to further improve the performance. Experimental results show that the proposed model achieves competitive performance with 1/3 of the parameters of the encoder, compared with the full-parameter model. 
\end{abstract}
\noindent\textbf{Index Terms}: parameter-efficient, sparsely-gated mixture-of-experts, Conformer, cross-layer weight-sharing

\section{Introduction}
Nowadays, transformers and their variants have been successfully applied to end-to-end (E2E) automatic speech recognition (ASR) \cite{vaswani2017attention,dong2018speech,gulati2020conformer}. Transformers usually use stacks of self-attention and feed-forward networks (FFNs) to build an encoder and a decoder \cite{vaswani2017attention}, and then use the attention mechanism to bridge the encoded acoustic features and the representations of text token sequences \cite{chan2016listen}. Lately, as a variant, conformers are developed to augment transformers with convolution by helping the model capture locality \cite{gulati2020conformer}. Combined with techniques such as relative positional representations \cite{shaw2018self,dai2019transformer} and Macaron-style half-step FFNs \cite{lu2019understanding}, conformers further improve the performance of transformers in ASR. 

Despite the promising performance, many works show the over-parameterization of transformers \cite{fan2019reducing,lan2019albert}, which leads the models to require much memory storage during training and inference, and hence limits the usage of the models on-device. To reduce the memory cost, some works share the parameters of one or several transformer blocks so that the total number of the parameters of the model is much reduced \cite{dehghani2018universal,lan2019albert,li2019end,zhao2020universal,komatsu2022non}. These models use one or a few transformer blocks to encode features in a recursive manner, thus the number of parameters is less than the original transformers with the same depth. However, because of the fewer free model parameters, the capacity of network is inevitably influenced and the performance degrades as a result.

To address this issue, we propose to share the sparsely-gated mixture-of-experts (MoE) to improve the capacity of cross-layer parameter-shared conformers, and not increase the computation in the meanwhile. Specifically, we first design the second FFN of a conformer block to be a sparsely-gated MoE module to improve the capacity and share the grouped conformer blocks in a cross-layer manner. The sparsely-gated MoE uses a dynamic routing mechanism to activate only one or a part of experts during training and inference, which keeps comparable computation with non-MoE models and scales the capacity of the model \cite{shazeer2017outrageously,riquelme2021scaling,you2021speechmoe,fedus2021switch}. Then, to help the parameter-shared conformer blocks to adapt the hidden representations at different levels, we propose to make each block have its own router so that the blocks can have flexible routing paths for different level representations. We also use individual normalization layers of the blocks to make them adaptable and to ensure the statistics consistent as well \cite{xue2021go}. Further, we use knowledge distillation \cite{hinton2015distilling,li2014learning} to help parameter-shared model imitate the full-parameter model. Experimental results on the public AISHELL-1 dataset demonstrate that the proposed parameter-efficient models can achieve competitive performance with $1/3$ of encoder parameters, compared with the full-parameter model.


\begin{figure*}[!t]\centering
	\includegraphics[width=2.0\columnwidth]{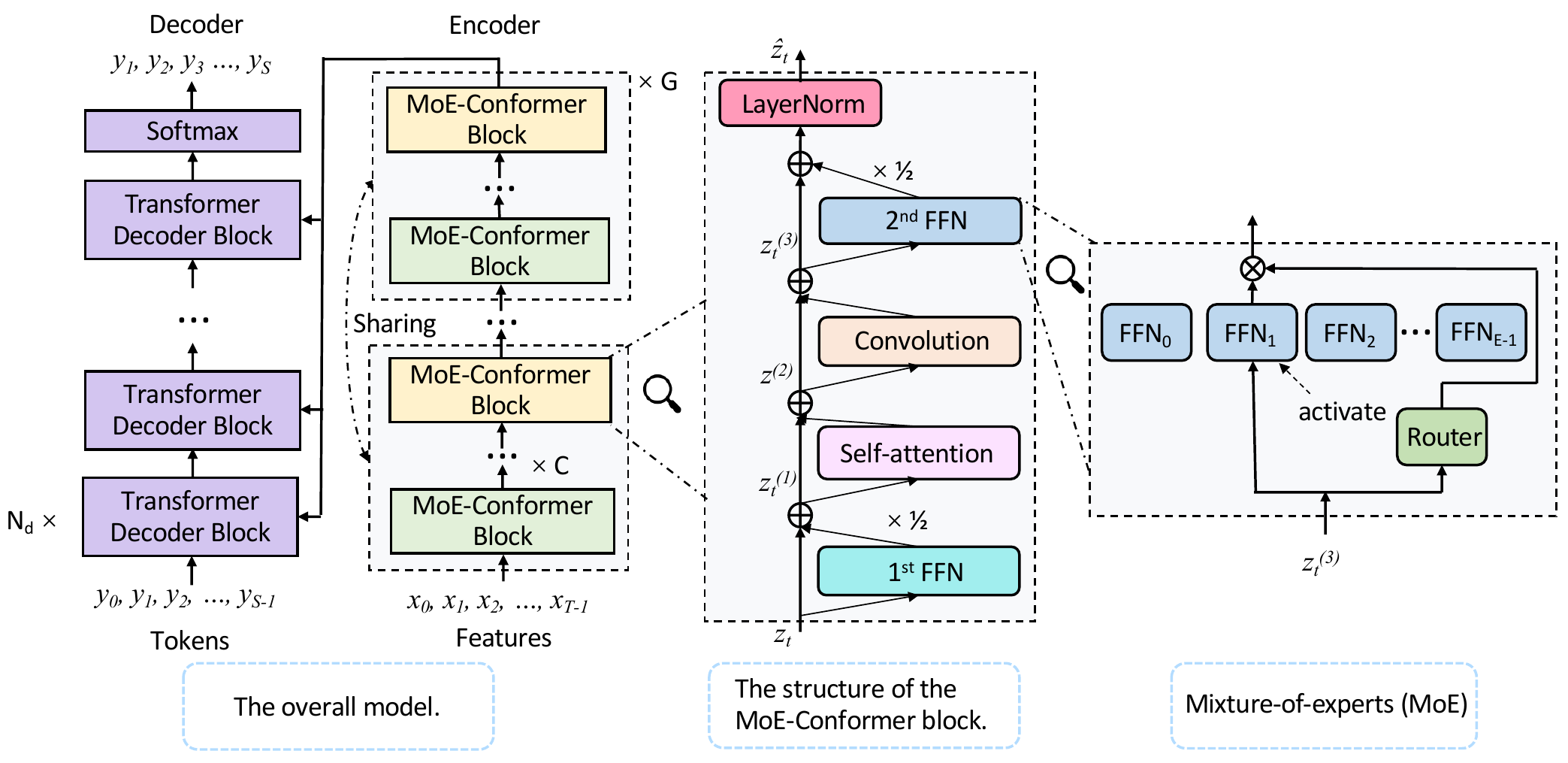}\vspace{-5pt}
	\caption{(\textbf{Left}) The overall architecture of the ASR system. The encoder consists of $G$ groups of $C$ consecutive MoE-conformer blocks. The parameters of the MoE-conformer blocks with the same color are shared among different groups, except the normalization modules and the routers of MoE module. The decoder consists of $\text{N}_\text{d}$ transformer decoder blocks. (\textbf{Mid.}) The structure of the MoE-conformer block consists of two feed-forward network (FFN) modules, a convolution module, and a multi-head self-attention (MHSA) module. The details of each module are referred to \cite{gulati2020conformer}. We novelly extend the second FFN module to the mixture-of-experts (MoE) recipe. (\textbf{Right}) The structure of the MoE module. The MoE module consists of several parallel FFN modules and a router. During forward propagation, the input is fed into one of the FFN modules which is activated by the router.}
	\vspace{-15pt}
	\label{fig:overall}
\end{figure*}

\vspace{-5pt}
\section{Background: Conformer-based Seq2Seq Models for ASR}
\vspace{-5pt}
\label{sec:bg}
As attention-based encoder-decoder (AED) models, transformers \cite{vaswani2017attention} use an encoder to capture the high-level representations from the acoustic features and a decoder to predict text sequences token-by-token with the attention mechanism. Formally, given an acoustic feature sequence $x = [x_0, \cdots, x_{T-1}]$ with length $T$ and a text token sequence $y = [y_0, \cdots, y_{S}]$ with length $S+1$, where $y_0$ and $y_S$ are the start-of-sentence symbol \texttt{<sos>} and the end-of-sentence symbol \texttt{<eos>}, the model $\textit{Trfm}$ predicts the probability of the text token:
\begin{equation}
    P(y_s | y_{<s}, x) = \textit{Trfm}(y_{<s}, x),
\end{equation}
where $y_{<s}$ is the prefix of $y_s$ in the text sequence, $1 \leq s \leq S$. The model is trained with maximum likelihood criterion:
\begin{align}
    L_\text{nll}(\theta) = -\frac{1}{S}\sum_{s=1}^{S} \log P(y_s | y_{<s}, x),
\end{align}
where $\theta$ is the parameters of the model $\textit{Trfm}$. The beam-search algorithm is used to find the most likely text token sequence during inference. The overall structure is shown in \autoref{fig:overall}.

Conformers \cite{gulati2020conformer} insert convolution layers into a transformer block to help the model capture locality of a sequence. With carefully designed fine-grained structures, including pre-norm \cite{xiong2020layer}, GLU \cite{dauphin2017language} and Swish \cite{ramachandran2017searching} activation functions, relative positional encodings \cite{shaw2018self,dai2019transformer}, conformers further improve the performance and stabilize the training. Our model chooses conformers as the basic structure of the encoder. The structure of the conformer block is shown in the middle part of \autoref{fig:overall}. The details of each module in a conformer block are referred to \cite{gulati2020conformer}.

\section{Sharing Sparsely-Gated Experts}
\label{sec:ssge}
\vspace{-5pt}
The core idea of the proposed parameter-efficient model is to reuse the conformer encoder blocks recursively to make the most of them. Crucially, the sparsely-gated MoE modules are used to improve the capacity of the modules without increasing computation. What’s equally important, the routers and the normalization layers are further designed in an individual way, so that they can be used as adapters to help the reused blocks adapt representations at different levels. The method can also be used in the other network structures, such as transformer decoder blocks and convolutional neural networks, as well as in other E2E ASR models, such as transducers \cite{graves2012sequence} and CTC \cite{graves2006connectionist}.

\vspace{-5pt}
\subsection{Parameter-Sharing for Conformers}
\vspace{-5pt}
The structure of a conformer, as shown in the middle part of \autoref{fig:overall}, consists of two FFN modules, a multi-head attention module, and a convolution module. All the modules use pre-norm style combination of the residual connection and the layer normalization. Besides, the FFN modules use Swish activation functions. The multi-head self-attention module uses relative positional encodings. The convolution module is a time-depth separable style convolutional block with GLU and Swish activation functions. More details of each module are referred to \cite{gulati2020conformer}. Here, formally, for the input representation $z_t$ at $t$ time-step, the computation of a conformer block is as follows:
\begin{equation}
    \begin{split}
    &z_t^{(1)} = z_t + \frac{1}{2}\text{\textit{FFN}}(z_t), \\
    &z_t^{(2)} = z_t^{(1)} + \text{\textit{MHSA}}(z_t^{(1)}), \\
    &z_t^{(3)} = z_t^{(2)} + \text{\textit{Conv}}(z_t^{(2)}), \\
    &\hat{z}_t = \text{\textit{LayerNorm}}(z_t^{(3)} + \frac{1}{2}\text{\textit{FFN}}^{(\text{\textit{MoE}})}(z_t^{(3)})),
    \end{split}
\end{equation}
where $\text{\textit{FFN}}$, $\text{\textit{MHSA}}$, $\text{\textit{Conv}}$ and $\text{\textit{FFN}}^{(\text{\textit{MoE}}})$ denote the first FFN module, the multi-head self-attention module, the convolution module, and the second FFN enhanced with MoE, respectively. $\textit{LayerNorm}$ denotes layer normalization \cite{ba2016layer}. The details of $\text{\textit{FFN}}^{(\text{\textit{MoE}}})$ are described in \autoref{sec:moe}

As shown in the left part of \autoref{fig:overall}, we share the parameters of different blocks. Specifically, $C$ consecutive conformer blocks are grouped and $G$ groups are stacked. For the conformer block at the same position in different groups, the parameters of each module are shared. It can be viewed as one group of conformer blocks are reused $G$ times, and the computation is implemented in a recursive iteration manner. Thus the model makes the most use of the parameters.

\vspace{-5pt}
\subsection{Dynamic Routing for Mixture of Experts}
\vspace{-5pt}
\label{sec:moe}
By parameter-sharing, the parameters of the encoder are much reduced. However, the capacity of the model is also reduced which influences the performance of the model negatively. So, to improve the model capacity but not increase the computation, we introduce sparsely-gated MoE \cite{shazeer2017outrageously,fedus2021switch} to the second FFN module, as shown in the right part of \autoref{fig:overall}. 

The sparsely-gated MoE mechanism consists of $E$ parallel experts and a router. The input $z_t^{(3)}$ is first fed into the router to select one of the experts\footnote{We use top-1 MoE to keep the number of the activated parameters the same with the non-MoE model in this paper.} and then is computed by the activated expert. The formal computation is as follows:
\begin{equation}\label{eq:moe}        
\begin{split}
g = [g_0, \cdots, g_{E-1}] &= {softmax}( {router}(z_t^{(3)}) ), \\
                   i^* &= \mathop{\arg\max}_{ 0 \leq i \leq E-1} g_i, \\
\text{\textit{FFN}}^{(\text{\textit{MoE}})}(z_t^{(3)})) &= g_{i^*} {\text{\textit{FFN}}}_{i^*}(z_t^{(3)}),
\end{split}
\end{equation}
where $\text{\textit{FFN}}_{i}$ denotes the $i$-th expert, $g_i$ denotes the gating value regarding to the $i$-th expert, and $i^*$ is the index of the selected expert. One may notice that, the procedure of MoE is actually similar to the attention mechanism: the input $z_t^{(3)}$ can be viewed as the query vector in the attention mechanism, and the gating scores $g$ can be viewed as the attention coefficients \cite{vaswani2017attention}. However, the attention procedure is in a ``hard'' way, namely, the non-maximum coefficients are all set to zero. 

In addition, to encourage all the experts to be used in balance, the load balancing loss \cite{fedus2021switch} is used as follows:
\begin{equation}
L_{\text{balance}} = E\sum_{i=0}^{E-1} f_i \bar{g}_i,
\end{equation}
where $f_i$ is the active frequency of the $i$-th expert in a batch, and $\bar{g}_i$ is the mean of the gating values computed for the $i$-th expert. Otherwise, Gaussian noises are added to the routers to make the expert selection various during training. 

With MoE, the parameters are extended so that the model capacity is increased. However, since only one FFN is activated actually, the computation is not increased.

\vspace{-5pt}
\subsection{Individual Routers and Normalization}
\vspace{-5pt}
To further improve the ability of the reused MoE modules, we propose to make each MoE module have its own router. The underneath thinking is to help the routing path achieve more flexibility. With this, the MoE modules in different MoE-conformer blocks can thus be adapted to different levels of representations. Furthermore, all normalization layers (including layer normalization and batch normalization) are built individually, thus, to maintain the statistics of the normalization layers corresponding to representations at different levels is consistent. And the scale and offset parameters in the normalization layers can be seen as parameter-efficient bias adapters \cite{zaken2021bitfit}.

\vspace{-5pt}
\subsection{Distilling Knowledge from Hidden Embedding}
\vspace{-5pt}
We use knowledge distillation \cite{li2014learning,hinton2015distilling} to transfer the knowledge from a full-parameter model to further improve the performance of the shared-parameter model. Specifically, we minimize the $L_2$ distance between the outputs of the shared-parameter encoder (student) and the full-parameter encoder (teacher):
\begin{equation}
    L_{kd} = \frac{1}{T} \sum_{t=0}^{T-1}|| h_t - h_t' ||,
\end{equation}
where $h_t$ denotes the output of the shared-parameter encoder and $h_t'$ denotes the output of the full-parameter encoder.

\vspace{-5pt}
\subsection{Learning}
\vspace{-5pt}
The model is learned by minimizing the overall loss:
\begin{equation}
    \label{eq:loss}
    L = L_\text{nll} + \alpha\frac{1}{C}\sum L_{\text{balance}} + \beta L_{kd},
\end{equation}
where $C$ is the number of MoE module (see \autoref{fig:overall}), $\alpha$ and $\beta$ are hyperparameters to balance the values of the losses. 

\vspace{-5pt}
\section{Relation to Prior Work}
\label{sec:rel}
\noindent\textbf{Conditional computation of mixture-of-experts}. MoE has been shown as an effective way to scale the capacity of neural networks without increasing computation \cite{shazeer2017outrageously,riquelme2021scaling,you2021speechmoe,fedus2021switch}. However, previous works aim to scale the model sizes to billions or trillions, which needs extremely much resources and model parallelization 
during training and inference. Different from these works, we aim to use MoE in a parameter-efficient way. We reuse the MoE modules to make the most use of them.

\noindent\textbf{Cross-layer weight sharing}. Cross-layer weight sharing is first used in transformer with adaptive computation time \cite{dehghani2018universal}. \cite{lan2019albert} uses this technique to reduce the parameters of BERT. \cite{li2019improving,zhao2020universal,komatsu2022non} use the similar techniques for ASR. However, directly sharing parameters may influence the capacity of the model negatively. To address this issue, we propose to use the MoE mechanism to improve the model capacity without increasing computation. Recently, \cite{xue2021go} shares the MoE module for ALBERT and ViT and applies the models to NLP and CV tasks. However, their work uses two experts, which increases computation cost. Otherwise, sharing the routers limits the capacity of the models. Different from their work, this paper focuses on more efficient architecture of Conformers \cite{gulati2020conformer} in ASR tasks. We use individual routers to help the model to have diverse routing paths at different levels. And we use group strategy to improve the model capacity in depth.

\begin{table}[!t]\centering
\caption{The overall character error rates on AISHELL-1. $N_{pe}$ denotes the total number of the parameters of the encoder. Dev. and Test denote the character error rates (CERs) on the development set and the test set, respectively.}
\begin{tabular}{|l||r|r|r|}
\hline
 \multicolumn{1}{|c||}{Model} & \multicolumn{1}{c|}{$N_{pe}$} & \multicolumn{1}{c|}{Dev.} & \multicolumn{1}{c|}{Test} \\ \hline\hline
\texttt{C12}                          & 21.58M                          & 4.46                     & 4.93                      \\ \hline\hline
\texttt{C2}                           & 3.74M                          & 5.86                     & 6.50                   \\ \hline
\texttt{C2-MoE4}                      & 6.89M                          & 5.77                     & 6.22                      \\ \hline
\texttt{C2-G6}                        & 3.74M                          & 5.18                     & 5.62                      \\ \hline
\texttt{C2-MoE4-G6}             & 6.95M                        & 4.67                     & 5.08                      \\ \hline
\texttt{C2-MoE4-G6-KD}          & 6.95M                        & \textbf{4.65}            & \textbf{5.03}             \\ \hline
\end{tabular}
\label{tab:aishell_overall}
\vspace{-10pt}
\end{table}
\vspace{-5pt}

\begin{table*}[t]
\caption{Ablation studies on AISHELL-1. ``n.'' denotes normalization modules, and ``r.'' denotes routers. ``indiv.'' means that the corresponding modules are not shared.}
\vspace{-10pt}
\begin{subtable}{0.5\linewidth}
\centering
\caption{w/ MoE vs. w/o MoE.}
\begin{tabular}{|l||r|r|r|}
\hline
 \multicolumn{1}{|c||}{Model}  & \multicolumn{1}{c|}{$N_{pe}$} & \multicolumn{1}{c|}{Dev.} & \multicolumn{1}{c|}{Test} \\ \hline\hline
\texttt{C1}      & 1.95M                         & 7.53                      & 8.41                      \\ \hline
\texttt{C1-MoE4} & 3.53M                         & 7.26                      & 8.05                      \\ \hline\hline
\texttt{C2}      & 3.74M                         & 5.86                      & 6.50                      \\ \hline
\texttt{C2-MoE4} \quad\quad\quad & 6.89M         & 5.77                      & 6.22                      \\ \hline
\end{tabular}
\label{tab:moe}

\vspace{12pt}

\caption{w/ parameter-sharing vs. w/o parameter-sharing.}
\begin{tabular}{|l||l|l|l|}
\hline
 \multicolumn{1}{|c||}{Model} & \multicolumn{1}{c|}{$N_{pe}$} & \multicolumn{1}{c|}{Dev.} & \multicolumn{1}{c|}{Test}  \\ \hline\hline
\texttt{C1}          & 1.95M                         & 7.53                      & 8.41                       \\ \hline
\texttt{C1-G12}      & 1.95M                         & 5.65                      & 6.07                       \\ \hline
\texttt{C1-MoE4-G12} & 3.59M                         & 5.01                      & 5.40                       \\ \hline\hline
\texttt{C2}          & 3.74M                         & 5.86                    & 6.50                         \\ \hline
\texttt{C2-G6}       & 3.74M                         & 5.18                    & 5.62                         \\ \hline
\texttt{C2-MoE4-G6}  & 6.95M                         & 4.67                      & 5.08                       \\ \hline
\end{tabular}
\label{tab:param_sharing}
\end{subtable} 
\begin{subtable}{0.5\linewidth}
\centering
\caption{Individual routers and normalization.}
\begin{tabular}{|l||r|r|r|}
\hline
 \multicolumn{1}{|c||}{Model}  & \multicolumn{1}{c|}{$N_{pe}$} & \multicolumn{1}{c|}{Dev.} & \multicolumn{1}{c|}{Test} \\ \hline\hline
\texttt{C1-MoE4-G12} (all shared)           & 3.53M                         & 6.39                      & 6.90                      \\ \hline
\texttt{C1-MoE4-G12} (indiv. n.)            & 3.58M                         & 5.19                      & 5.57                      \\ \hline
\texttt{C1-MoE4-G12} (indiv. n. \& r.)      & 3.59M                         & 5.01                      & 5.40                      \\ \hline\hline
\texttt{C2-MoE4-G6} (all shared)            & 6.89M                         & 5.60                      & 6.00                      \\ \hline
\texttt{C2-MoE4-G6} (indiv. n.)             & 6.94M                         & 4.72                      & 5.21                      \\ \hline
\texttt{C2-MoE4-G6} (indiv. n. \& r.)       & 6.95M                         & 4.67                      & 5.08                      \\ \hline
\end{tabular}
\label{tab:aishell_ind}

\vspace{5pt}

\caption{Knowledge distillation from hidden embeddings.}
\begin{tabular}{|l||r|r|r|}
\hline
 \multicolumn{1}{|c||}{Model} & \multicolumn{1}{c|}{$N_{pe}$} & \multicolumn{1}{c|}{Dev.} & \multicolumn{1}{c|}{Test}  \\ \hline\hline
\texttt{C12} (teacher)     & 21.58M                        & 4.46                      & 4.93              \\ \hline\hline                  
\texttt{C1-MoE4-G12}       & 3.53M                         & 5.01                      & 5.40              \\ \hline
\texttt{C1-MoE4-G12-KD}    & 3.53M                         & 4.99                      & 5.43              \\ \hline\hline
\texttt{C2-MoE4-G6}        & 6.95M                          & 4.67                      & 5.08              \\ \hline
\texttt{C2-MoE4-G6-KD} \quad\quad\quad\quad\quad  & 6.95M   & 4.65                      & 5.03              \\ \hline
\end{tabular}
\label{tab:aishell_kd}
\end{subtable}
\vspace{-15pt}
\end{table*}


\vspace{-5pt}
\section{Experiments}
\vspace{-5pt}
\label{sec:exp}

\subsection{Experimental Setup}
\vspace{-5pt}
We conduct experiments on a publicly available Chinese Mandarin AISHELL-1\footnote{https://www.openslr.org/33/} dataset \cite{bu2017aishell}, which includes about 150
hours of speech for training, about 18 hours of speech for development, and about 10 hours speech for test.

For all the experiments, we use 80-dimension Mel-filter bank features (FBANK) as the
input, which are extracted every $10$\,ms with $25$\,ms window. We use global cmvn as feature normalization. Speed perturbation with factors of 0.9, 1.0 and 1.1 is used as audio augmentation \cite{ko2015audio}. All the feature processing is employed with Kaldi toolkit \cite{povey2011kaldi}. We use 4235 Chinese characters as the vocabulary, including \texttt{<sos>} and \texttt{<eos>}.

We use a 2-layer CNN as a subsampling module. Each layer is a $3\times3$ convolutional layer with $32$ output channels, and the stride is $2$. Thus, the frame rate is subsampled to $25$ Hz. For the encoder, we set the dimension of an MoE-Conformer module to $256$, the number of heads of MHSA to $4$, the kernal size of Conv to $15$. The intermediate dimension of an FFN module is $1024$. We use $4$ experts for the second FFN in an MoE-Conformer module. We compare effects of the different number of MoE-Conformer modules and groups, i.e., $C$ and $G$ in \autoref{fig:overall}. For the decoder, we use the transformer structure. To control experimental variables, we fix the number of the decoder blocks to $4$. The dimension of the decoder module is also $256$ and the intermediate dimension of the FFN in the decoder module is $1024$. We set dropout rate to $0.1$ and use SpecAugmentation \cite{park2019specaugment} and time stretch \cite{nguyen2020improving} to avoid overfitting. The values of $\alpha$ and $\beta$ in \autoref{eq:loss} are set to $0.01$ and $0.005$, respectively. The standard deviation of Gaussian noise for the MoE gate is set to $0.1$ in training. CTC loss is also used to improve the alignment with weight 0.2. The learning rate schedule is inverse square root with 4000 warm-up steps. All models are trained for 80 epochs with 8 GPU cards. One batch includes 32000 frames. We use PyTorch \cite{paszke2019pytorch} and FastMoE \cite{he2021fastmoe} for implementation.

\begin{figure}[!t]\centering
    \vspace{-5pt}
	\includegraphics[width=0.8\columnwidth]{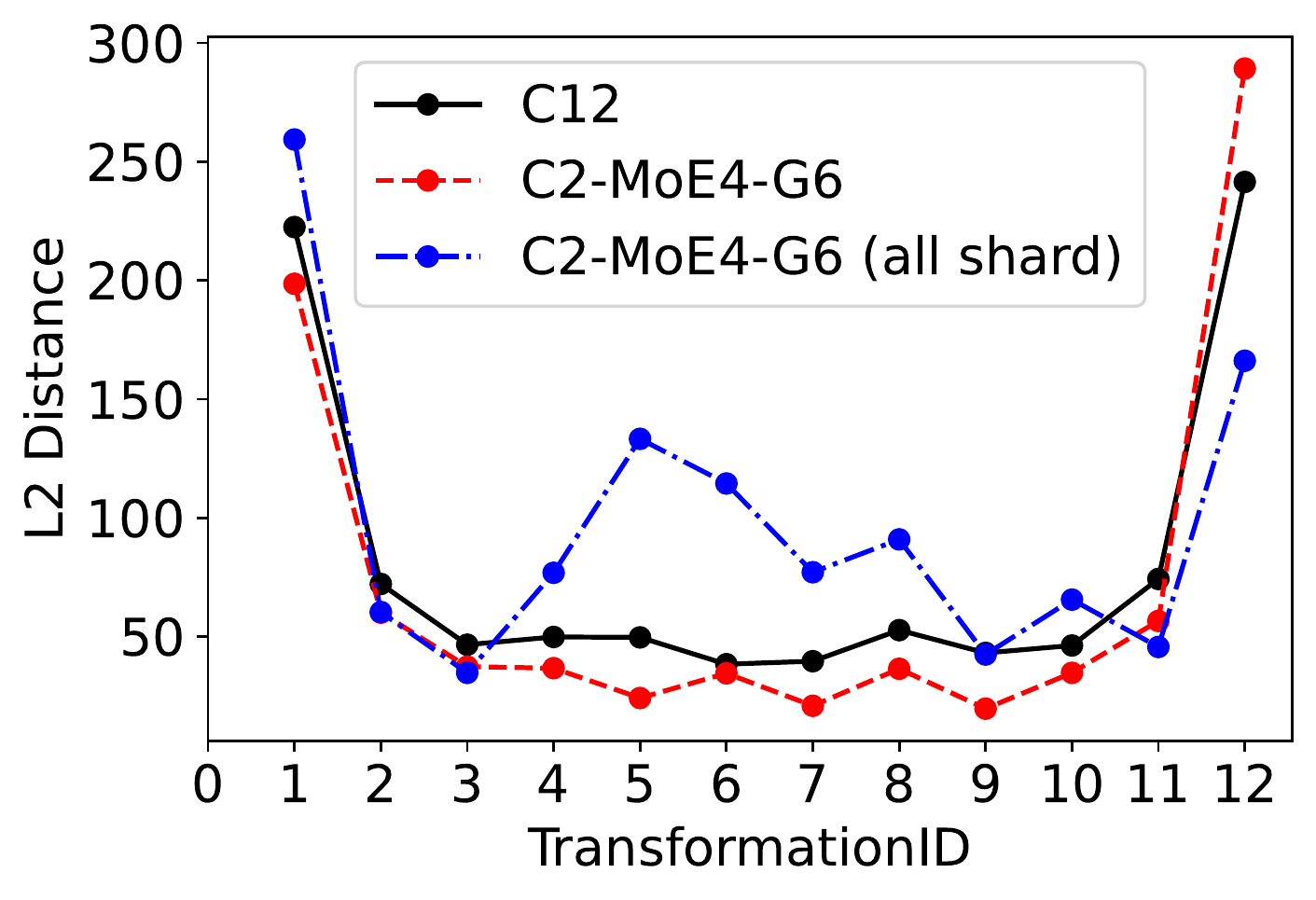}\vspace{-5pt}
	\caption{L2 distances between the input and output for each transformation.}
	\vspace{-15pt}
	\label{fig:curve}
\end{figure}

\vspace{-5pt}
\subsection{Results and Analysis}
\vspace{-5pt}

\textbf{Overall}. \autoref{tab:aishell_overall} shows the overall performance of our model. \texttt{C} denotes the number of conformer blocks and \texttt{G} denotes the number of groups (see \autoref{fig:overall}). \texttt{MoE4} and \texttt{KD} mean whether to use MoE and knowledge distillation, respectively. We can see that with the proposed methods, \texttt{C2-MoE4-G6-KD} achieves competitive performance with $1/3$ parameters of the encoder, compared with the full-parameter model \texttt{C12}. Directly reducing the number of blocks to 2 hurts the performance of the model (\texttt{C2}). MoE improves the performance of the models and does not increase the activated parameters.

\noindent\textbf{w/ MoE vs. w/o MoE}. We compare the shallow encoders with MoE and the ones without MoE in \autoref{tab:moe}. We can see that MoE improves the capacity so that the performance of \texttt{C1-MoE4} is better than \texttt{C1} and the performance of \texttt{C2-MoE4} is better than \texttt{C2}. And with more conformer blocks, the model can perform better but with more parameters (\texttt{C2} vs. \texttt{C1} and \texttt{C2-MoE4} vs. \texttt{C1-MoE4}).

\noindent\textbf{Recursive iteration}. \autoref{tab:param_sharing} shows that more recursive iterations make better performance with the same number of parameters. Specifically, for \texttt{C1-G12}, the 12 groups of blocks are shared, which can be seen as the group is computed recursively, and the performance is much better than the non-shared model \texttt{C1}. Similarly, \texttt{C2-G6} performs better than \texttt{C2} with the same number of parameters. \texttt{C2-G6}, which has more blocks in one group than \texttt{C1-G12}, performs better than \texttt{C1} with the same computation iteration. MoE improves capacity and performance of \texttt{C1-G12} and \texttt{C2-G6}.

\noindent\textbf{Individual routers and normalization}. \autoref{tab:aishell_ind} compares the effect of the individual routers and normalization. We can see that if the routers and the normalization modules are all shared, the performance of the parameter-shared model is hurt heavily. Keeping each conformer block having its own normalization module at each group can make the modules have proper statistics so that the performance is better. And the individual routers make the MoE module be able to select proper experts at different levels. Thus, the model with individual routers and normalization can achieve better performance.

\noindent\textbf{Knowledge distillation from hidden embeddings}. We further use knowledge distillation to make the parameter-sharing model imitate the full-parameter model (\autoref{tab:aishell_kd}). We can see that with knowledge distillation, the performance of the parameter-shared model is further improved for \texttt{C2-MoE4-G6} model. However, the improvement is not very significant for \texttt{C1-MoE4-G12}. This is probably because, \texttt{C1} model has far less  conformer blocks when comparing to \texttt{C12} model ($1$ vs. $12$), then such a big divergence  influences the effect of knowledge distillation \cite{cho2019efficacy} between them.

\noindent\textbf{L2 distances between the input and the output}. \autoref{fig:curve} shows the L2 distances of the input and output of each transformation for an example utterance. We can see that \texttt{C2-MoE4-G6} shows a similar behavior with the full-parameter model \texttt{C12}. Whereas the curve of the all-shared model is oscillating. This shows that the individual routers and normalization have an effect on stabilizing network parameters.

\vspace{-5pt}
\section{Conclusions and Future Works}
\label{sec:conc}
This paper explores sharing the sparsely-gated mixture-of-experts (MoE) to build a parameter-efficient conformer model for speech recognition. Specifically, we first use MoE to extend the capacity of a conformer block. Then, we share the parameters of the grouped conformer blocks so that the parameters are much reduced compared with the full-parameter model. To ensure the representations adapt at different levels, we make the routers of MoE and normalization modules individual. Moreover, we use knowledge distillation to further improve the performance. The experimental results demonstrate that the proposed model can achieve competitive performance with about 1/3 of the parameters of the encoder, compared with the full-parameter model. In the future, we will extend the proposed method to more large-scale datasets and other ASR models, such as transducers and CTC.

\newpage
\bibliographystyle{IEEEtran}

\bibliography{mybib}

\end{document}